\documentclass[twocolumn]{aastex631}

\usepackage{amsmath}
\usepackage{graphicx}
\usepackage{array}
\usepackage{xspace}
\usepackage{multirow}
\usepackage{appendix}
\usepackage{fontawesome5}
\usepackage{booktabs, multirow}
\usepackage{ulem}
\usepackage{float}

\newcommand{\angstrom}{\mbox{\normalfont\AA \ }}
\newcommand{\angstromnospace}{\mbox{\normalfont\AA }}

\newcommand{\Otwo}{[OII]\xspace}
\newcommand{\Hbeta}{H$\beta$\xspace}
\newcommand{\Othreea}{[OIII]$\lambda\, 4959$\AA\xspace}
\newcommand{\Othreeb}{[OIII]$\lambda\, 5007$\AA\xspace}

\usepackage{comment}
\usepackage{hyperref}

\begin{document} 

\title{Quasars acting as Strong Lenses Found in DESI DR1}

\author{E.~McArthur}
\affiliation{Physics Department, Stanford University, Stanford, CA 93405, USA}
\affiliation{Kavli Institute for Particle Astrophysics and Cosmology, Stanford University, Menlo Park, CA 94305, USA}
\affiliation{Department of Astronomy, The Ohio State University, 140 W 18th Ave, Columbus, OH 43210}

\author[0000-0001-7051-497X]{M.~Millon}
\affiliation{Institute for Astronomy, ETH Z\"{u}rich, Wolfgang-Pauli-Strasse 27, CH-8093 Z\"{u}rich, Switzerland}
\affiliation{SLAC National Accelerator Laboratory, 2575 Sand Hill Road, Menlo Park, CA 94025, USA}
\affiliation{Kavli Institute for Particle Astrophysics and Cosmology, Stanford University, Menlo Park, CA 94305, USA}

\author[0000-0003-2284-8603]{M.~Powell}
\affiliation{SLAC National Accelerator Laboratory, 2575 Sand Hill Road, Menlo Park, CA 94025, USA}

\author[0000-0003-2229-011X]{R.~H.~Wechsler}
\affiliation{Kavli Institute for Particle Astrophysics and Cosmology, Stanford University, Menlo Park, CA 94305, USA}
\affiliation{Physics Department, Stanford University, Stanford, CA 93405, USA}
\affiliation{SLAC National Accelerator Laboratory, 2575 Sand Hill Road, Menlo Park, CA 94025, USA}

\author[0000-0003-0230-6436]{Z.~Pan}
\affiliation{Kavli Institute for Astronomy and Astrophysics at Peking University, PKU, 5 Yiheyuan Road, Haidian District, Beijing 100871, P.R. China}

\author[0000-0002-2949-2155]{M.~Siudek}
\affiliation{Institut de F\'{i}sica d’Altes Energies (IFAE), The Barcelona Institute of Science and Technology, Edifici Cn, Campus UAB, 08193, Bellaterra (Barcelona), Spain}
\affiliation{Institute of Space Sciences, ICE-CSIC, Campus UAB, Carrer de Can Magrans s/n, 08913 Bellaterra, Barcelona, Spain}
\affiliation{Instituto de Astrof\'{\i}sica de Canarias, C/ V\'{\i}a L\'{a}ctea, s/n, E-38205 La Laguna, Tenerife, Spain}

\author{J.~Spiller}
\affiliation{Institute for Astronomy, ETH Z\"{u}rich, Wolfgang-Pauli-Strasse 27, CH-8093 Z\"{u}rich, Switzerland}

\author{J.~Aguilar}
\affiliation{Lawrence Berkeley National Laboratory, 1 Cyclotron Road, Berkeley, CA 94720, USA}

\author[0000-0001-6098-7247]{S.~Ahlen}
\affiliation{Department of Physics, Boston University, 590 Commonwealth Avenue, Boston, MA 02215 USA}

\author[0000-0003-2923-1585]{A.~Anand}
\affiliation{Lawrence Berkeley National Laboratory, 1 Cyclotron Road, Berkeley, CA 94720, USA}

\author[0000-0001-5537-4710]{S.~BenZvi}
\affiliation{Department of Physics \& Astronomy, University of Rochester, 206 Bausch and Lomb Hall, P.O. Box 270171, Rochester, NY 14627-0171, USA}

\author[0000-0001-9712-0006]{D.~Bianchi}
\affiliation{Dipartimento di Fisica ``Aldo Pontremoli'', Universit\`a degli Studi di Milano, Via Celoria 16, I-20133 Milano, Italy}
\affiliation{INAF-Osservatorio Astronomico di Brera, Via Brera 28, 20122 Milano, Italy}

\author{D.~Brooks}
\affiliation{Department of Physics \& Astronomy, University College London, Gower Street, London, WC1E 6BT, UK}

\author{T.~Claybaugh}
\affiliation{Lawrence Berkeley National Laboratory, 1 Cyclotron Road, Berkeley, CA 94720, USA}

\author[0000-0002-2169-0595]{A.~Cuceu}
\affiliation{Lawrence Berkeley National Laboratory, 1 Cyclotron Road, Berkeley, CA 94720, USA}

\author[0000-0002-1769-1640]{A.~de la Macorra}
\affiliation{Instituto de F\'{\i}sica, Universidad Nacional Aut\'{o}noma de M\'{e}xico,  Circuito de la Investigaci\'{o}n Cient\'{\i}fica, Ciudad Universitaria, Cd. de M\'{e}xico  C.~P.~04510,  M\'{e}xico}

\author[0000-0002-4928-4003]{Arjun~Dey}
\affiliation{NSF NOIRLab, 950 N. Cherry Ave., Tucson, AZ 85719, USA}

\author{P.~Doel}
\affiliation{Department of Physics \& Astronomy, University College London, Gower Street, London, WC1E 6BT, UK}

\author[0000-0002-3033-7312]{A.~Font-Ribera}
\affiliation{Institut de F\'{i}sica d’Altes Energies (IFAE), The Barcelona Institute of Science and Technology, Edifici Cn, Campus UAB, 08193, Bellaterra (Barcelona), Spain}

\author[0000-0002-2890-3725]{J.~E.~Forero-Romero}
\affiliation{Departamento de F\'isica, Universidad de los Andes, Cra. 1 No. 18A-10, Edificio Ip, CP 111711, Bogot\'a, Colombia}
\affiliation{Observatorio Astron\'omico, Universidad de los Andes, Cra. 1 No. 18A-10, Edificio H, CP 111711 Bogot\'a, Colombia}

\author[0000-0001-9632-0815]{E.~Gaztañaga}
\affiliation{Institut d'Estudis Espacials de Catalunya (IEEC), c/ Esteve Terradas 1, Edifici RDIT, Campus PMT-UPC, 08860 Castelldefels, Spain}
\affiliation{Institute of Cosmology and Gravitation, University of Portsmouth, Dennis Sciama Building, Portsmouth, PO1 3FX, UK}
\affiliation{Institute of Space Sciences, ICE-CSIC, Campus UAB, Carrer de Can Magrans s/n, 08913 Bellaterra, Barcelona, Spain}

\author[0000-0003-3142-233X]{S.~Gontcho A Gontcho}
\affiliation{Lawrence Berkeley National Laboratory, 1 Cyclotron Road, Berkeley, CA 94720, USA}
\affiliation{University of Virginia, Department of Astronomy, Charlottesville, VA 22904, USA}

\author{G.~Gutierrez}
\affiliation{Fermi National Accelerator Laboratory, PO Box 500, Batavia, IL 60510, USA}

\author[0000-0002-9136-9609]{H.~K.~Herrera-Alcantar}
\affiliation{Institut d'Astrophysique de Paris. 98 bis boulevard Arago. 75014 Paris, France}
\affiliation{IRFU, CEA, Universit\'{e} Paris-Saclay, F-91191 Gif-sur-Yvette, France}

\author[0000-0002-6550-2023]{K.~Honscheid}
\affiliation{Center for Cosmology and AstroParticle Physics, The Ohio State University, 191 West Woodruff Avenue, Columbus, OH 43210, USA}
\affiliation{Department of Physics, The Ohio State University, 191 West Woodruff Avenue, Columbus, OH 43210, USA}
\affiliation{The Ohio State University, Columbus, 43210 OH, USA}

\author[0000-0002-6024-466X]{M.~Ishak}
\affiliation{Department of Physics, The University of Texas at Dallas, 800 W. Campbell Rd., Richardson, TX 75080, USA}

\author[0000-0003-0201-5241]{R.~Joyce}
\affiliation{NSF NOIRLab, 950 N. Cherry Ave., Tucson, AZ 85719, USA}

\author[0000-0002-0000-2394]{S.~Juneau}
\affiliation{NSF NOIRLab, 950 N. Cherry Ave., Tucson, AZ 85719, USA}

\author[0000-0002-8828-5463]{D.~Kirkby}
\affiliation{Department of Physics and Astronomy, University of California, Irvine, 92697, USA}

\author[0000-0003-3510-7134]{T.~Kisner}
\affiliation{Lawrence Berkeley National Laboratory, 1 Cyclotron Road, Berkeley, CA 94720, USA}

\author[0000-0001-6356-7424]{A.~Kremin}
\affiliation{Lawrence Berkeley National Laboratory, 1 Cyclotron Road, Berkeley, CA 94720, USA}

\author{O.~Lahav}
\affiliation{Department of Physics \& Astronomy, University College London, Gower Street, London, WC1E 6BT, UK}

\author[0000-0002-6731-9329]{C.~Lamman}
\affiliation{The Ohio State University, Columbus, 43210 OH, USA}

\author[0000-0003-1838-8528]{M.~Landriau}
\affiliation{Lawrence Berkeley National Laboratory, 1 Cyclotron Road, Berkeley, CA 94720, USA}

\author[0000-0001-7178-8868]{L.~Le~Guillou}
\affiliation{Sorbonne Universit\'{e}, CNRS/IN2P3, Laboratoire de Physique Nucl\'{e}aire et de Hautes Energies (LPNHE), FR-75005 Paris, France}

\author[0000-0003-4962-8934]{M.~Manera}
\affiliation{Departament de F\'{i}sica, Serra H\'{u}nter, Universitat Aut\`{o}noma de Barcelona, 08193 Bellaterra (Barcelona), Spain}
\affiliation{Institut de F\'{i}sica d’Altes Energies (IFAE), The Barcelona Institute of Science and Technology, Edifici Cn, Campus UAB, 08193, Bellaterra (Barcelona), Spain}

\author[0000-0002-1125-7384]{A.~Meisner}
\affiliation{NSF NOIRLab, 950 N. Cherry Ave., Tucson, AZ 85719, USA}

\author{R.~Miquel}
\affiliation{Instituci\'{o} Catalana de Recerca i Estudis Avan\c{c}ats, Passeig de Llu\'{\i}s Companys, 23, 08010 Barcelona, Spain}
\affiliation{Institut de F\'{i}sica d’Altes Energies (IFAE), The Barcelona Institute of Science and Technology, Edifici Cn, Campus UAB, 08193, Bellaterra (Barcelona), Spain}

\author[0000-0001-9070-3102]{S.~Nadathur}
\affiliation{Institute of Cosmology and Gravitation, University of Portsmouth, Dennis Sciama Building, Portsmouth, PO1 3FX, UK}

\author[0000-0003-3188-784X]{N.~Palanque-Delabrouille}
\affiliation{IRFU, CEA, Universit\'{e} Paris-Saclay, F-91191 Gif-sur-Yvette, France}
\affiliation{Lawrence Berkeley National Laboratory, 1 Cyclotron Road, Berkeley, CA 94720, USA}

\author[0000-0002-0644-5727]{W.~J.~Percival}
\affiliation{Department of Physics and Astronomy, University of Waterloo, 200 University Ave W, Waterloo, ON N2L 3G1, Canada}
\affiliation{Perimeter Institute for Theoretical Physics, 31 Caroline St. North, Waterloo, ON N2L 2Y5, Canada}
\affiliation{Waterloo Centre for Astrophysics, University of Waterloo, 200 University Ave W, Waterloo, ON N2L 3G1, Canada}

\author{C.~Poppett}
\affiliation{Lawrence Berkeley National Laboratory, 1 Cyclotron Road, Berkeley, CA 94720, USA}
\affiliation{Space Sciences Laboratory, University of California, Berkeley, 7 Gauss Way, Berkeley, CA  94720, USA}
\affiliation{University of California, Berkeley, 110 Sproul Hall \#5800 Berkeley, CA 94720, USA}

\author[0000-0001-7145-8674]{F.~Prada}
\affiliation{Instituto de Astrof\'{i}sica de Andaluc\'{i}a (CSIC), Glorieta de la Astronom\'{i}a, s/n, E-18008 Granada, Spain}

\author[0000-0001-6979-0125]{I.~P\'erez-R\`afols}
\affiliation{Departament de F\'isica, EEBE, Universitat Polit\`ecnica de Catalunya, c/Eduard Maristany 10, 08930 Barcelona, Spain}

\author{G.~Rossi}
\affiliation{Department of Physics and Astronomy, Sejong University, 209 Neungdong-ro, Gwangjin-gu, Seoul 05006, Republic of Korea}

\author[0000-0002-9646-8198]{E.~Sanchez}
\affiliation{CIEMAT, Avenida Complutense 40, E-28040 Madrid, Spain}

\author{D.~Schlegel}
\affiliation{Lawrence Berkeley National Laboratory, 1 Cyclotron Road, Berkeley, CA 94720, USA}

\author{M.~Schubnell}
\affiliation{Department of Physics, University of Michigan, 450 Church Street, Ann Arbor, MI 48109, USA}
\affiliation{University of Michigan, 500 S. State Street, Ann Arbor, MI 48109, USA}

\author[0000-0002-6588-3508]{H.~Seo}
\affiliation{Department of Physics \& Astronomy, Ohio University, 139 University Terrace, Athens, OH 45701, USA}

\author[0000-0002-3461-0320]{J.~Silber}
\affiliation{Lawrence Berkeley National Laboratory, 1 Cyclotron Road, Berkeley, CA 94720, USA}

\author{D.~Sprayberry}
\affiliation{NSF NOIRLab, 950 N. Cherry Ave., Tucson, AZ 85719, USA}

\author[0000-0003-1704-0781]{G.~Tarl\'{e}}
\affiliation{University of Michigan, 500 S. State Street, Ann Arbor, MI 48109, USA}

\author{B.~A.~Weaver}
\affiliation{NSF NOIRLab, 950 N. Cherry Ave., Tucson, AZ 85719, USA}

\author[0000-0001-5381-4372]{R.~Zhou}
\affiliation{Lawrence Berkeley National Laboratory, 1 Cyclotron Road, Berkeley, CA 94720, USA}

\author[0000-0002-6684-3997]{H.~Zou}
\affiliation{National Astronomical Observatories, Chinese Academy of Sciences, A20 Datun Road, Chaoyang District, Beijing, 100101, P.~R.~China}

\collaboration{500}{(DESI Collaboration)}

\begin{abstract}
Quasars acting as strong gravitational lenses offer a rare opportunity to probe the redshift evolution of scaling relations between supermassive black holes and their host galaxies, particularly the $M_{\rm BH}$–$M_{\rm host}$ relation. Using these powerful probes, the mass of the host galaxy can be precisely inferred from the Einstein radius $\theta_E$. Using 812,118 quasars from DESI DR1 ($0.03 \leq z \leq 1.8$), we searched for quasars lensing higher-redshift galaxies by identifying background emission-line features in their spectra. To detect these rare systems, we trained a convolutional neural network (CNN) on mock lenses constructed from real DESI spectra of quasars and emission-line galaxies (ELGs), achieving a high classification performance (AUC = 0.99). We also trained a regression network to estimate the redshift of the background ELG. Applying this pipeline, we identified 7 high-quality (Grade A) lens candidates, each exhibiting a strong \Otwo doublet at a higher redshift than the foreground quasar; four candidates additionally show \Hbeta, \Othreea, and \Othreeb emission. These results significantly expand the sample of quasar lens candidates beyond the 12 identified and 3 confirmed in previous work, and demonstrate the potential for scalable, data-driven discovery of quasars as strong lenses in upcoming spectroscopic surveys.

\end{abstract}


\section{Introduction} \label{sec:intro} 

Supermassive black holes (SMBHs) show strong correlations with key properties of their host galaxies, including stellar mass ($M_*$), total mass ($M_{\rm host}$), and stellar velocity dispersion ($\sigma_*$) \citep{2000ApJ...539L...9F, 2000ApJ...539L..13G, Magorrian, Gultekin, Ferrarese, McConnell, Powell_2022}. These relationships suggest a coevolutionary link between galaxies and their central black holes, but the physical origin of this connection---whether driven by feedback, merger histories, or other processes---remains uncertain.

To better understand this coevolution, it is critical to trace the evolution of SMBH–host galaxy scaling relations, such as $M_{\rm BH}$–$\sigma_*$ and $M_{\rm BH}$–$M_{\rm host}$, across cosmic time \citep{2013ARA&A..51..511K}. 

Two main scenarios have been proposed to explain the observed SMBH–host galaxy scaling relations. In one, Active Galactic Nuclei (AGN) feedback regulates star formation and black hole accretion by coupling quasar-driven outflows to the surrounding medium. This mechanism can reproduce the observed correlations in simulations \citep{2005Natur.435..629S, 2008ApJ...676...33D}, although the strength of the effect remains debated. In the alternative hierarchical assembly scenario, correlations emerge statistically as galaxies and black holes grow via mergers \citep{2007ApJ...671.1098P, Hirschmann, Jahnke}. Discriminating between these scenarios requires precise measurements of how the slope, normalization, and scatter of these relations evolve with redshift \citep{2009Knud, 2013Schramm}.

Observational results to date remain mixed. Pre-JWST studies did not find a strong evolution of redshift in the $M_{\rm BH}$ -$\sigma_*$ relation \citep{Sexton_2019, Suh, YZhang}, while others reported increasing $M_{\rm BH}/M_*$ ratios with redshift \citep{2021MNRAS.501..269D, Ding_2020}. These trends may suggest early SMBH growth, followed by later bulge build-up via processes such as minor mergers or disk instabilities. Conversely, studies showing little evolution in $M_{\rm BH}$–$\sigma_*$ and $M_{\rm BH}$–$M_{\rm host}$ \citep{Schulze_2014, Sexton_2019} imply that previously observed offsets may stem from selection effects rather than intrinsic change \citep{2008ASPC..399..419W, Treu_2007}. Hydrodynamical simulations have attempted to model this evolution, but show divergent predictions depending on their black hole and feedback prescriptions \citep{Habouzit}.

With the advent of the James Webb Space Telescope \citep[JWST,][]{2023PASP..135f8001G}, observations of high-redshift AGN and improved QSO–host decompositions have become possible. Many recent studies report that SMBHs in $z > 4$ AGN and QSOs are overmassive relative to their hosts \citep{Ding, Kocevski, Harikane, Maiolino, Yue, Stone, Mezcua_2023, Mezcua_2024}, typically lying $1$–$2$ dex above the local relation. Although the interpretation of this offset is debated due to potential measurement biases \citep{Pacucci_2023, Volonteri, Li, Sun}, these results are consistent with an intrinsic evolution in $M_{\rm BH}/M_*$.

Yet, even with these advances, measuring host galaxy properties at high redshift remains difficult. Quasar light often dominates photometry, complicating decomposition, and stellar mass estimates are sensitive to assumptions about the initial mass function (IMF) and star formation history \citep{Siudek_2024, 2007ApJ...670..249L}. A powerful alternative arises in rare systems where the quasar host acts as a gravitational lens on a background galaxy. In such cases, the total projected mass, typically on galaxy scales where baryons dominate over dark matter, can be inferred from the Einstein radius ($\theta_E$). This offers a direct measurement of $M_{\rm host}$ that avoids photometric modeling uncertainties and better correlated with $M_{\rm BH}$ \citep{Millon_2023, 2013ARA&A..51..511K}.

The first such system was discovered spectroscopically in the Sloan Digital Sky Survey \citep[SDSS;][]{SDSS, Courbin2010}. Three additional systems were identified in SDSS DR7 and confirmed with follow-up imaging from the Hubble Space Telescope \citep{Courbin_2012}. \citet{Meyer_2019} identified 12 additional candidates in SDSS-III BOSS, including quasars lensing both emission-line galaxies (ELGs) and Lyman-$\alpha$ emitters (LAEs). These systems await space-based confirmation, but collectively represent the current sample of known or suspected QSO lenses.

However, the rarity of these systems remains a limiting factor. A statistically significant sample of QSO lenses spanning several redshift bins is essential to constrain the evolution of $M_{\rm BH}$–$M_{\rm host}$. While future imaging surveys could uncover hundreds more \citep{Taak2020}, all current detections have come from spectroscopic searches. Previous studies using the Dark Energy Spectroscopic Instrument \citep[DESI][]{desicollaboration2016desiexperimentiiinstrument, desicollaboration2016desiexperimentisciencetargeting, miller2023opticalcorrectordarkenergy, Poppett} have primarily focused on identifying galaxy-galaxy lenses or lensed quasars through imaging \citep{Storfer, Dawes_and_storfer, Huang2025}, rather than detecting background emission lines in fiber spectra as done in SLACS and BELLS \citep{Bolton2006, Shu2015, Brownstein2012, Shu2016}. The relatively large fiber diameters of SDSS ($3\arcsec$) and BOSS ($2\arcsec$) mean that the presence of two distinct redshift components in a single spectrum typically signals a close alignment, consistent with strong lensing.

These spectroscopic searches relied on template fitting or principal component analysis (PCA) to subtract the foreground object before searching for residual emission lines \citep{Meyer_2019}. While effective for galaxy--galaxy lenses, this approach struggles for QSO lenses because of the spectral complexity and variability of quasars. No single template or PCA basis can simultaneously model all QSO emission features with high accuracy, leading to high false positive rates unless key features are masked \citep{Meyer_2019}.

In this work, we overcome these limitations by developing a fast, scalable, machine learning–based pipeline to identify quasar–galaxy lens candidates in DESI Year 1 data. These rare systems enable direct measurements of host galaxy mass via the Einstein radius of the background lensed source, providing a model-independent constraint on $M_{\rm host}$. Our method addresses key limitations of prior approaches by training on real DESI spectra to learn the full diversity of quasar features and is designed to scale efficiently to the millions of quasar spectra expected from DESI.
Our convolutional neural network is trained on mock lenses constructed from real DESI QSO and ELG spectra, capturing realistic noise, diversity, and blending effects. A complementary regression network estimates the redshift of the background source. This method enables robust identification of lenses across the large and heterogeneous data set of DESI and significantly expands the sample of QSO lens candidates, paving the way for future studies of the evolution of the SMBH host with unprecedented statistical power.

This paper is organized as follows. In Section \ref{sec:data} we describe how we construct our training samples and testing samples. In Section \ref{sec:methods} we describe how we construct the neural network for a binary lens/non-lens classification and to find the source redshift of the QSO-lens candidates.  We then conclude with a description of the results in Section \ref{Results} as well as future plans and considerations in Section \ref{Discussion}.

\section{Data} \label{sec:data}
The Dark Energy Spectroscopic Instrument (DESI) is a fiber-fed spectrograph designed to measure the expansion history of the universe to investigate the effects of dark energy for eight years \citep{Schlafly_DESI, DESI_Collaboration_2022}. 

The first data release (DR1) of DESI provided spectra for approximately 19 million objects including galaxies, quasars and stars to provide the first \citep{desicollaboration2025datarelease1dark}. The results of DR1 also set new constraints on dark energy \citep{Adame_2025, Abdul_Karim_2025}. A measurement of the redshift for both galaxies and quasars is possible because of a wavelength coverage between 3600\angstrom and 9800\angstrom and a high spectral resolution between 2000 to 5500.
 
 In the first data release of DESI, $\sim$1.8 million quasar spectra are provided across a wide redshift range. For information about the target selection of the DESI QSO's, see \cite{Chaussidon_2023} and \cite{Raichoor_2023}. We select 812,118 quasars using the HEALPixel-based \citep{HEALPIX_REFERENCE} redshift catalog from the main DESI DR1 survey using the ``dark time" program; this is to select spectra that are not influenced by increased noise in the blue camera due to the moon (``bright time" program). 
 We use information from Redrock, which is part of the spectroscopic processing pipeline used by DESI \citep{bailey_redrock, Redshiftestimation_abhijeet}. The output from Redrock includes the QSO's TARGETID, redshift ($z$), and the associated redshift error \citep{Guy_2023}. Based on these criteria, we select only objects with OBJTYPE = TGT and ZCAT\_PRIMARY = 1 to ensure the exclusion of sky targets and spectra that are not designated as primary spectra and do not have redshift fitting failures. Lastly, we select objects with ZWARN = 0 and SPECTYPE = QSO -- filtering out objects that could have been targeted as QSO but have a different spectral classification. This approach ensures that the redshifts are accurate and that we are only training off of quasar spectra where the redshift calculation shows no indications of error. 

After selecting the quasars, we used the FastSpec catalog to construct our ELG sample; this step is crucial to constructing mock lenses. The catalog is based on FastSpecFit\footnote{More information on this can be found at \url{https://fastspecfit.readthedocs.io/en/latest/index.html}} \citep{Moustakas2023}, a lightweight pipeline that provides information on DESI objects, including emission-line fluxes, redshifts, classifications, and more. FastSpecFit uses templates to fit specific parameters and spectral models and to construct a noise-free spectrum of the object. We first select the ELGs in the same way as the QSOs and with the same redshift range as the QSOs; however, the SPECTYPE = GALAXY. From this filtering, we select 16,500 Emission Line Galaxies (ELGs) but only select the ELGs from the main survey where $\text{OII\_3726\_FLUX} > 2 \times 10^{-17}~\mathrm{erg\,cm^{-2}\,s^{-1}}$, which allows us to select ELGs that have emission above the noise level of the data. 

In this analysis we restrict both ELGs and QSOs to the redshift range $0.03 \leq z \leq 1.8$. At $z \geq 1.8$ the Lyman-$\alpha$ forest begins to enter the observed spectrum, complicating line identification. Although \Otwo\ falls out of the DESI wavelength range at $z \simeq 1.6$, we expect our neural network to generalize to quasars at $1.6 \leq z \leq 1.8$. A systematic search for other emission lines detectable at higher redshifts—such as Lyman-$\alpha$—is left for future work.

\section{Methods} \label{sec:methods}
To create a successful model for identifying quasars acting as strong lenses, it is essential to train the model on a labeled dataset that enables it to distinguish between the spectral properties of a quasar acting as a lens and one that does not. This training process allows the model to generalize to a blind sample, where the true positive cases are unknown.

Using the QSO and the ELG spectra observed by DESI, we train a convolutional neural network (CNN) on mock lens systems (positive samples) and unlensed QSO spectra (negative samples) using a 10\% to 90\% ratio, respectively. The reason for this unbalanced training set is that we expect lenses to be rarer than non-lenses. However, we cannot reproduce this extremely low lens rate in our training rate because the CNN would not see enough positive examples. This split is a trade-off to minimize the rate of false positive while still showing enough positive samples to the CNN. One approach is to have unbalance classes but with a large augmentation of the lensing rate in the training set, see \citep{Lanusse, Savary2022, Petrillo} for a discussion. We train using real observed data to include all the possible astrophysical and instrumental nuisances in the training set. However, the same data cannot be used for training and for searching real lenses. Therefore, our data set is split and used in two phases. Phase 1 uses 47\% of the 812,118 objects, and the rest are used for Phase 2 then we use a 70/30 split for both Phase samples training and validation. Specifically: 

\textbf{Phase 1}:
    \begin{itemize}
        \item \textbf{Training Sample}: 70\% of 384,873 QSOs of the Phase 1 Training Sample were used to train the classification and redshift-finding networks.
        \item \textbf{Validation Sample}: remaining 30\% of 384,873 QSOs from the Phase 1 Training Sample were set aside to validate the performance of the models during training.
        \item \textbf{Blind Sample}: The blind sample consists of 427,245 QSOs that were not used during the training, validation, and testing of our method. We search for real lenses in this data set after training our methods.
        \item \textbf{Test Sample}: For testing, we used 3,170 QSOs from the Phase 1 blind sample and generated mock lens systems for 10\% of them. This test sample is used to assert the performance of the network after the optimization of the hyper-parameters.

    \end{itemize}
    
\textbf{Phase 2}:
    In Phase 2, we invert the training and the blind sample, and repeat the same procedure.
    \begin{itemize}
        \item \textbf{Training Sample}: We train the CNN on 70 \% of the 427,245 QSOs from the Phase 1 Blind Sample.
        \item \textbf{Validation Sample}: During validation, we use 30\% of the 427,245.
        \item \textbf{Blind Sample}: The blind sample consists of 384,873 QSOs that were not used during training, validation, and testing. 
        \item \textbf{Test Sample}: For testing in Phase 2, we used 3,547 QSOs from the Phase 2 blind sample, which were separate from the training and validation subsets. 10\% of the QSOs are constructed lensed systems.
    \end{itemize}

We note that with this approach, the real lenses contained in DESI can be mislabeled during training. However, we do not expect that these mislabeled samples would impact the performance of the network, as the lensing occurrence is extremely low.

\subsection{Quasar Strong Lens Classification} \label{CNN}
\subsubsection{Training Set Construction and Mock Lensed Systems}
One could use simulated spectra for both the QSO and ELG to generate the training set, but here we aim to preserve characteristics that are inherent to DESI spectra, whether due to the instrument, observational conditions, or due to the objects themselves. QSO spectra and ELG \Otwo profiles are very diverse for various reasons (e.g., broadening, luminosity; \citealt{Brodzeller_2022,lan2024desiemissionlinegalaxies}), so we use the observations obtained from DR1 to directly train our neural network on real data and therefore match by construction the survey's population properties of the QSOs and ELGs. 

\begin{figure*}[!hbt]
    \centering
    \includegraphics[width=\textwidth]{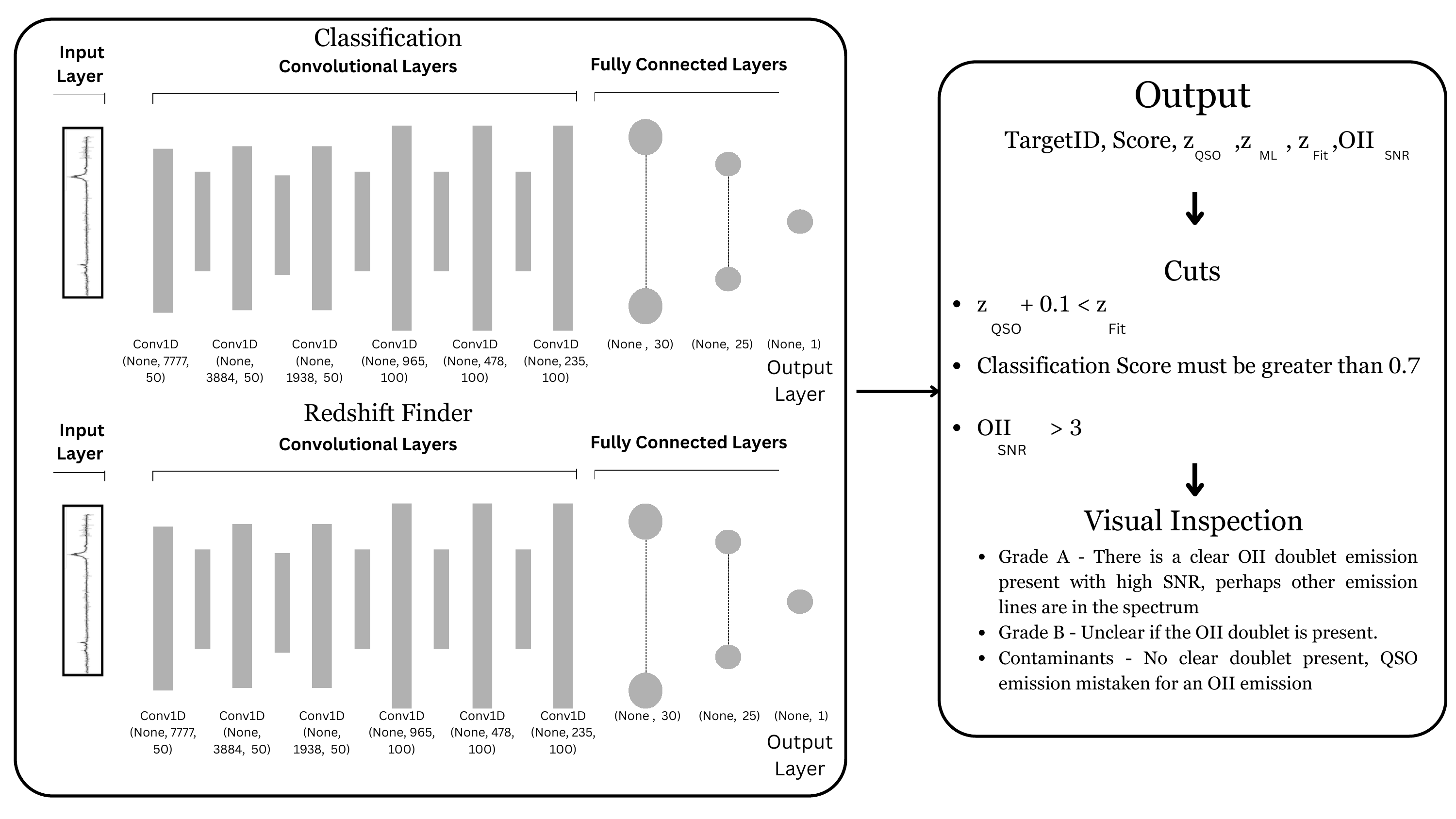}
    \caption{Diagram of the full pipeline illustrating the sequence of steps used to identify quasar (QSO) lens systems in the DESI DR1 dataset. The process integrates convolutional neural networks for classification, redshift-based selection criteria, and visual inspection to refine candidates based on score thresholds, emission-line strengths, and model fits. Futher explanation for the cuts and inspection can be found in section \ref{RedshiftTraining}.}
    \label{fig:Pipeline}
\end{figure*}

However, QSOs acting as strong lenses are extremely rare. In SDSS, there were 12 candidates out of 297301 quasars \citep{Meyer_2019} and only 1 of them is in the DESI DR1 dataset within our redshift range. Consequently, our positive samples, i.e., the QSO lenses, need to be constructed by adding the flux of a real QSO and of higher redshift ELG to simulate the spectrum of a gravitational lens and obtain a sizable training set containing a sufficient number of positive and negative samples. Equation \ref{model} describes how the fluxes of the QSO and ELG are summed to obtain a ``mock" lens: 
\begin{equation}
\label{model}
\text{$f_\mathrm{L}$} = \text{$f_\mathrm{QSO}$}+ \mu \cdot \text{$f_\mathrm{ELG}$}, 
\end{equation}
where f$_{QSO}$ is the QSO flux and f$_{ELG}$ is the noiseless spectra of the higher redshift ELG randomly drawn from FastSpec \citep{Moustakas2023} multiplied by a lens magnification factor ($\mu$). The value of $\mu$ is randomly drawn from a normal distribution with mean $\mu=4$ and standard deviation $\sigma_\mu=2$, to emulate the typical total magnification factor expected for lenses at the galaxy scale \citep{Oguri2010}. The size of the DESI fiber is $1.5\arcsec$ in diameter, which is larger than the typical image separation for QSO lenses, so we expect the fiber to capture the flux of both the foreground QSO and the magnified flux of the background ELG. Therefore, the detection of such blended spectra will have a very high probability of being a lens, as demonstrated in \cite{Courbin_2012}, whose candidates were confirmed by HST high-resolution imaging with a success rate of 75\%. The complete pipeline steps can be found in Figure \ref{fig:Pipeline}; here we explain each step in detail.

\label{Pipeline}

\subsubsection{CNN classifier training and architecture}
\label{Architecture and CNN}
CNN Models are deep learning algorithms particularly well suited for processing images and spectroscopic data, especially when noise or resolution is considered \citep{2015Natur.521..436L, NIPS2012_c399862d}. The core component of a CNN is the convolutional layer, which is designed to extract local features from the input data.

A convolutional layer applies learnable filters to the input, creating feature maps that identify patterns. These maps are processed through multiple convolutional layers to capture complex features and then passed to fully connected layers for interpretation and prediction.  In our application, we expect the different filters to capture the different emission lines that characterize QSOs and ELGs spectra. Our CNN architecture consists of six convolutional layers: three layers with 50 filters and three layers with 100 filters. This is followed by two fully connected layers, these layers use nodes instead of filters, the nodes are computational units that output a value based on inputs from other nodes or the convolutional layers. The first layer has 30 nodes and the second layer has 25 nodes as demonstrated in the left panel of Figure \ref{fig:Pipeline}. The network outputs a score between 0 and 1. During training, we set the threshold value to an arbitrary value of 0.5. Any prediction score at or above 0.5 is considered a lens by the neural network (this threshold is later adjusted based on the desired balance between completeness and purity). During the training, we use the Adam optimizer \citep{kingma2017adammethodstochasticoptimization} with an exponential learning rate decay, where the learning rate decreases every 500 steps by a factor of 0.95. This process was completed using TensorFlow \citep{tensorflow2015-whitepaper} and scikit-learn \citep{scikit-learn} was used for splitting the training set, generating the confusion matrix and metrics. For the loss function during training we use Binary Cross-entropy described in the Equation \ref{eq:BCE}, where $\hat{y}_i$ is the predicted value for the sample, $y_i$ is the true value for the sample, and N is the number of samples:

\begin{equation}
\label{eq:BCE}
\mathcal{L}_{\text{BCE}} = -\frac{1}{N} \sum_{i=1}^{N} 
\Bigl[
y_i \log(\hat{y}_i) + (1 - y_i) \log(1 - \hat{y}_i)
\Bigr].
\end{equation}

We then use the Phase 1 training sample and the Phase 2 training sample to take advantage of two CNNs, a CNN trained on the Phase 1 training sample whose weights are applied to the Phase 1 blind sample and a separate CNN trained on the Phase 2 training sample and applied to the Phase 2 blind sample. 

After training, the model is tested on the test sample to determine the classification performance. The threshold is adjusted to maximize the $F_1$ score which uses the true positive (TP), false positive (FP), and false negative (FN) rates as in Equation \ref{equation 2}:
\begin{equation}
\label{equation 2}
F_1 = \frac{2 \cdot \text{TP}}{2 \cdot \text{TP} + \text{FP} + \text{FN}} .
\end{equation}

\begin{figure}[hbt!]
    \centering
    \includegraphics[width=0.47\textwidth]{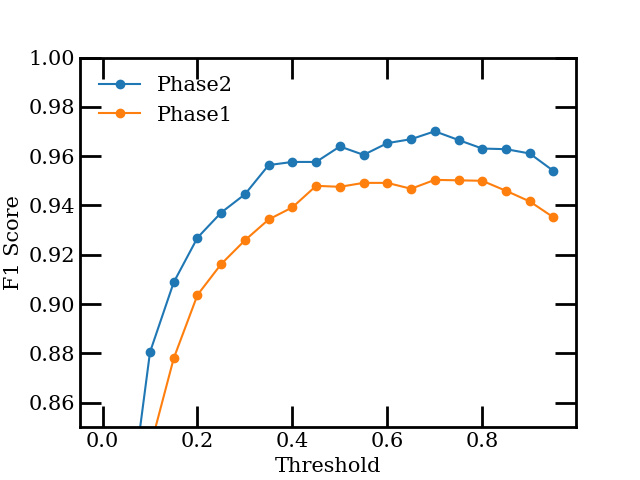}
    \caption{$F_1$ Score, calculated for different threshold values for both Phase 1 and Phase 2. Our final adopted threshold maximizes the F1 score with an adopted value of 0.70 for Phase 1 and Phase 2.}
    \label{fig:F1ScorevThreshold}
\end{figure}

The F1 score is often used alongside accuracy because accuracy only measures the proportion of correctly labeled samples out of the total, while the F1 score accounts for how well the classifier balances precision and recall - how many true positive candidates it correctly identified (true positives) relative to the number of false positives and false negative candidates. In Figure \ref{fig:F1ScorevThreshold}, we show the $F_1$ score at each threshold value. The highest $F_1$ score corresponds to a threshold value of 0.7 for both phases. We adopt a value when the classifier is applied to the blind sample.

Finally, we apply each network to a blind sample, which consists of observed quasars that it has not seen to obtain the first list of candidate lenses.  

\subsection{Source Redshift Estimation} \label{High $z$}
The CNN for classification does not provide information on the redshift of the lensed source contained in the quasar spectrum. To address this issue and obtain the redshift of the source galaxy, we adopted two different methods and compared their performance. The first is built upon DESI's redshift finding pipeline, Redrock, while the second is built on a third CNN with a similar architecture as the classifier described in Section \ref{Architecture and CNN}.

\subsubsection{Redrock} 
 Redrock is designed to find the redshift of an observed spectrum by comparing the principle component analysis templates (PCA template) of stars, galaxies, and QSOs to the observed spectrum and by performing a grid search on a redshift range until the chi-squared $\chi^2(x)$ is minimized.

In our case, the foreground quasar's redshift is usually robustly determined by Redrock, as it is usually much brighter than the background object. However, it remains unclear whether Redrock can reliably recover the redshift of background source galaxies. To test this, we first use Redrock to fit the spectrum of the foreground quasar and subtract the best-fit model from the observed data. We then apply Redrock again to the residual spectrum to measure the redshift of the background ELG.

The Redrock output provides redshift solutions for each source, ranked by their reduced $\chi^2$, with the first solution being the most probable. As a result, we use the first solution to predict the source redshift ($z_{pred}$). We input both Phase 1 and Phase 2 test samples into Redrock. It is important to note that we only use the mock lensed systems from the test sample for this purpose, consisting of 575 lensed systems across both phase test samples. This ensures a consistent and unbiased comparison of the redshift predictions generated by both methods to compare Redrock's performance with the Redshift Finding CNN described in the next section.

\subsubsection{Redshift Finding CNN} \label{RedshiftTraining}
We also implement a regression model that predicts the source redshift solution ($z_\mathrm{ELG}$). This model employs the same architecture as the classification CNN, consisting of convolutional layers and a fully connected layer, but instead of the sigmoid activation function outputting a score between 0 and 1, it outputs the redshift value of the source through a mean squared error (MSE) loss function. The Mean Squared Error (MSE) loss function is defined as:

\begin{equation}
\mathcal{L}_{\text{MSE}} = \frac{1}{N} \sum_{i=1}^{N} \bigl(y_i - \hat{y}_i\bigr)^{2}
\end{equation}

We train the regression model only on mock lens systems generated following Equation \ref{model}, using the 326,689 QSOs from the Phase 1 training sample described in Section \ref{CNN} to create mock lens systems paired randomly with a higher redshift ELG.

Our performance metrics during training are the mean absolute error (MAE) and the MSE, computed by comparing the predicted redshift of the mock lens to the true redshift of the source. Specifically, the MSE measures the average squared difference between the actual and predicted values, penalizing larger errors more heavily. The MAE quantifies the average magnitude of errors in a set of predictions, providing a more interpretable measure of the accuracy of the prediction. 

After the network is fully trained, we add an additional step to refine the accuracy of the predicted redshift from the network ($z_\mathrm{ML}$). Starting from the redshift predicted by CNN, we fit a double Gaussian model with a fixed separation of 2.7\angstrom and a fixed amplitude ratio of 1:1.3 \citep{Kriek} to the \Otwo doublet in a redshift range of $\Delta z=0.1$ around $z_\mathrm{ML}$. As a result of this fit, we obtain a refined estimate of the redshift $z_\mathrm{fit}$ as well as an estimate of the signal-to-noise ratio (SNR) of the \Otwo line. Next, we input the Phase 1 Test Set and the Phase 2 Test Set explained in Section \ref{sec:data} into the full pipeline as shown in Figure \ref{fig:Pipeline}.
This includes inputting the spectrum into both the classifier and the regression network and obtaining the outputted results (left panel Figure \ref{fig:Pipeline}). Next we apply additional cuts to the calculated signal-to-noise ratio (SNR) of the \Otwo emission line to keep only high-quality candidates. We keep only candidates with an SNR $\geq$ 3. We further stipulate that the redshift of the QSO must be less than the ELG and the separation between the ELG and QSO's redshift must be greater than or equal to $\Delta z=0.1$ (right panel of Figure \ref{fig:Pipeline}). A smaller redshift separation has a very low chance of being a lens and this cut ensures that the QSO and the ELG are not physically associated, for example, in a merging system.


\section{Results} 
\label{Results}

\subsection{Classifier Performance}

The metric performance of the CNN classifier, when applied to the Phase 1 and Phase 2 training and validation sets, is presented in Table \ref{tab:CNNResults}. While the primary goal is to minimize the loss function, which quantifies how well the network predictions match the true values, Table \ref{tab:CNNResults} also reports the following performance metrics:

\begin{itemize}
    \item \textbf{Accuracy} – The ratio of correct predictions to the total number of predictions.
    \item \textbf{F1 Score} – A metric that balances precision and recall by computing an average that gives more influence to lower values. This ensures that if either precision or recall is low, the overall score remains low, making it a fair measure of a model’s performance in minimizing both false positives and false negatives.  
    \item \textbf{AUC} – The area under the Receiver Operating Characteristic (ROC) curve, which plots the true positive rate against the false positive rate across different threshold settings.  
\end{itemize}   

\begin{table}[!htp]
\centering
\begin{tabular}{lrrrr}
\toprule
& Loss & Accuracy & F1 Score & AUC \\
\midrule
\multicolumn{5}{c}{Training} \\
Phase 1 & 0.0217 & 0.994 & 0.8957 & 0.9951 \\
Phase 2 & 0.0386 & 0.989 & 0.8601 & 0.9900 \\
\midrule
\multicolumn{5}{c}{Validation} \\
Phase 1 & 0.0254 & 0.9931 & 0.8913 & 0.9923 \\
Phase 2 & 0.0338 & 0.9900 & 0.8767 & 0.9941 \\
\bottomrule
\end{tabular}
\caption{Phase 1 and Phase 2 training and validation results obtained for each metric.}
\label{tab:CNNResults}
\end{table}

\begin{figure}
    \centering
    \includegraphics[width=0.47\textwidth]{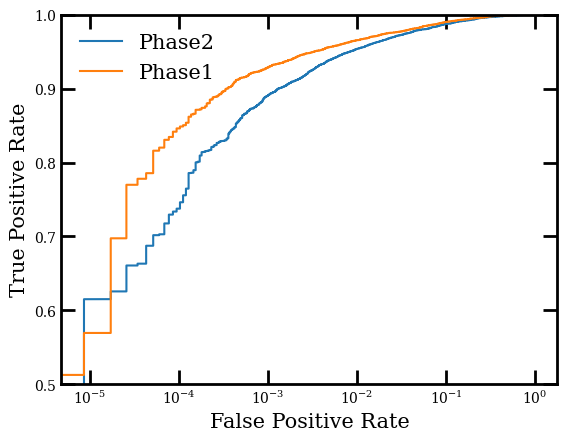}
    \caption{True Positive Rate vs False Positive Rate computed on the validation sample.}
    \label{fig:AUC}
\end{figure}

\label{Redrock}
\begin{figure*}[hbt!]
    \centering
    \includegraphics[width=0.8\textwidth]{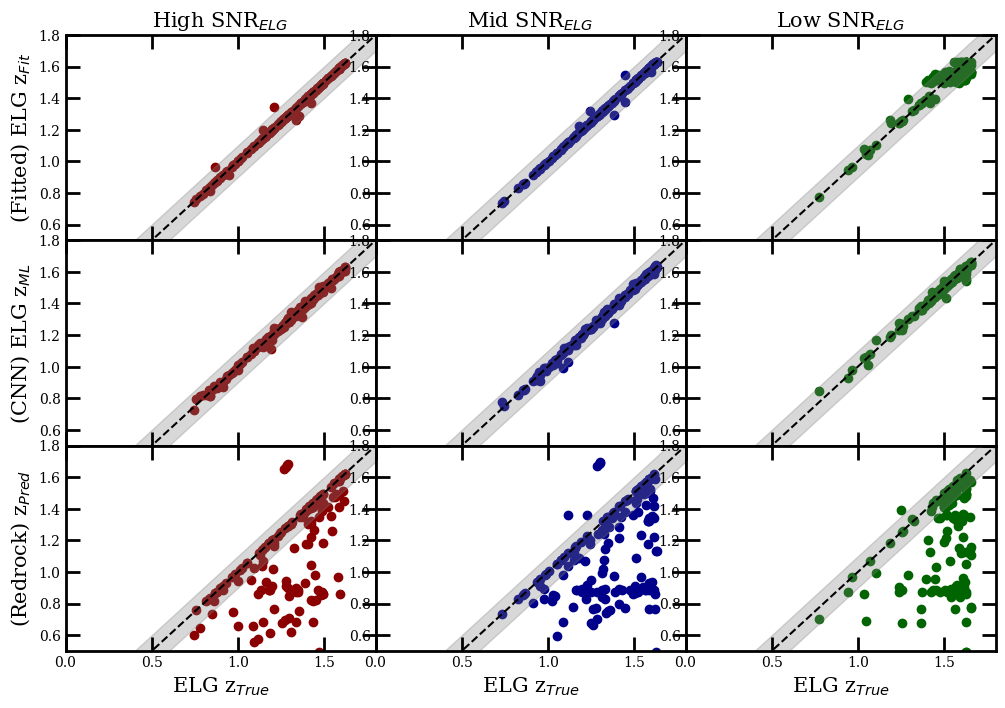}
    \caption{True redshifts of the background object $z_{ELG}$ versus the recovered redshifts $z_{\rm True}$ grouped by the SNR of the \Otwo \ line. The shaded region corresponds to a redshift error $\Delta z =0.1$ from the true redshift value.}
    \label{fig:SNR_Plot}
\end{figure*}

Figure \ref{fig:AUC} shows the ROC curve, which is how our network performs at each true positive rate for both phases. 

\subsection{Redshift Finder performances} \label{Model Selection}

\begin{table}[!htp]\centering
\begin{tabular}{lrrrrrrrrrrrrrrrrrrrrrrrrrr}\toprule
&Training Set & \\\midrule
Loss: &MSE: &MAE: & \\
1.95e-4 &1.95e-4 &0.006 & \\
&Validation Set & \\
Validation Loss &Validation MSE: &Validation MAE: & \\
8.49e-4 &8.49e-4 &8.2e-4 & \\
\bottomrule
\end{tabular}
\caption{Values for each metric used during training of the redshift finder.}
\label{tab: Regression Table}
\end{table}

After training the redshift finder, we report the MSE and MAE in Table \ref{tab: Regression Table}. While this result is encouraging, we also compare our method's performance with that of Redrock.

On our test sample, we rank each object by the SNR of the higher redshift ELG to see how both the CNN and Redrock perform within a given range of signal-to-noise of the \Otwo emission feature. We split the sample into 3 groups using their percentile for both phases: low SNR (3 $\leq$ SNR $<$ 7.52), mid SNR (7.52 $\leq$ SNR $<$ 16.63), high SNR ( SNR $\geq$ 16.63). 

Within the low SNR regime, we expect that the flux of the quasar in the foreground dominates the spectrum, and the features from the ELG would be difficult to discern by both the CNN and Redrock as opposed to the high SNR regime, where a strong \Otwo emission would be observed. To quantify the performance of Redrock, the CNN redshift finder, and the Gaussian fit to the \Otwo as described in Section \ref{RedshiftTraining}; we calculate the percentage of test samples recovered within $\Delta z = 0.1$ of the real redshift value (marked by the shaded region in Figure \ref{fig:SNR_Plot}). Within the high-SNR range, 100\% of the source redshift is recovered within $\Delta z = 0.1$ by the CNN, 99.48\% after the Gaussian fit, and 51.04\% for Redrock. In the mid-SNR region, we obtain 99.48\% for the CNN and 100\% after the Gaussian fit, compared to  37.70\% from Redrock, and lastly, for the low SNR, we obtain 100.00\% , 96.88\%, and 29.17\% for the CNN, Gaussian fit, and Redrock respectively. 


In summary, across all SNR regimes, the CNN redshift finder combined with a Gaussian fit recovers the background ELG redshift more accurately than Redrock. The infrared channel contains numerous skylines and noise residuals, even after masking, which makes the CNN perform better than the standard Redrock solution in this wavelength range. While the Gaussian fit yields nearly exact results in the mid-SNR regime, it struggles at very low SNR, where the pure CNN approach performs better. We still adopt the CNN + Gaussian fit method because the CNN provides an excellent initial estimate, which the Gaussian fit then refines to achieve higher precision in most of the higher SNR regimes.

\subsection{Application to the blind samples}

With a fully trained neural network we now apply the weights to both Phase 1 and Phase 2 blind samples. Using a threshold score of 0.7, the CNN predicts 494 candidates out of the 812,118 QSOs in both phases. We visually inspect the remaining candidates and grade them either A or B. Grade A candidates denote high-priority candidates that have a high SNR \Otwo emission line, with a clear doublet feature, and other lines visible in the spectrum such as \Hbeta and \Othreea,  if covered by the DESI spectral range. Grade B candidates at their fitted redshift ($z_{fit}$) are ambiguous, in which unclear whether the detected feature is the \Otwo doublet. Contaminants are objects that the classifier selected, but the redshift finder does not converge on an OII doublet. In some cases, the source redshift for these objects is the same as the QSO or based solely on a QSO emission line.
This process identifies 7 grade A candidates, whose properties are detailed in Table \ref{tab:Results}. The grade B candidates can be found in Table  \ref{Tab:GradeB} .

\begin{deluxetable*}{l r r r r r r r}
\tablecaption{QSO--ELG candidates found in DESI DR1 from the full pipeline.\label{tab:Results}}
\tablehead{
\colhead{Name} &
\colhead{TargetID} &
\colhead{$z_{\rm QSO}$} &
\colhead{Score} &
\colhead{$z_{\rm ML}$} &
\colhead{$z_{\rm fit}$} &
\colhead{$\chi^2$} &
\colhead{SNR [OII]}
}
\startdata
DESI J257.5186+07.4378 & 39627969840291993 & 0.5190 & 0.8017 & 0.7661 & 0.7775 $\pm$ 0.0102 & 1.1019 & 6.95 \\
DESI J186.8602+43.8706 & 39633144625761309 & 0.5089 & 0.9192 & 0.8175 & 0.8185 $\pm$ 0.0003 & 1.2217 & 7.16 \\
DESI J172.4216+33.1854 & 39632945656367381 & 0.7718 & 0.9999 & 1.2033 & 1.1972 $\pm$ 0.0004 & 1.1895 & 8.21 \\
DESI J215.0329+01.8025 & 39627830614557718 & 0.5616 & 0.9689 & 0.7401 & 0.8080 $\pm$ 0.0003 & 1.3988 & 9.53 \\
DESI J194.0101-03.6566 & 39627697403467094 & 0.5083 & 0.7482 & 1.2580 & 1.2986 $\pm$ 0.0008 & 1.3484 & 14.91 \\
DESI J007.8207+12.5949 & 39628084772604358 & 0.5343 & 0.9365 & 1.1983 & 1.2060 $\pm$ 0.0004 & 1.0459 & 8.17 \\
DESI J140.0528-02.3751$^{\dagger}$ & 39627726679704831 & 0.6578 & 0.8917 & 0.8581 & 0.7694 $\pm$ 0.0005 & 1.1811 & 7.99 \\
\enddata
\tablecomments{Column descriptions: $z_{\rm QSO}$ — redshift of the foreground QSO from DESI’s Redrock pipeline; Score — classification score (0–1); $z_{\rm fit}$ — fitted redshift from a Gaussian fit over a $\Delta z = 0.1$ window around $z_{\rm ML}$; $\chi^2$ — reduced goodness of fit for $z_{\rm fit}$; SNR — signal-to-noise ratio from the flux/noise integral in a 10\,\AA\ window around [O\,II]. The $\dagger$ indicates the candidate already found in SDSS by \citet{Meyer_2019}.}
\end{deluxetable*}

\begin{figure*}

    \centering
    \includegraphics[width=\textwidth]{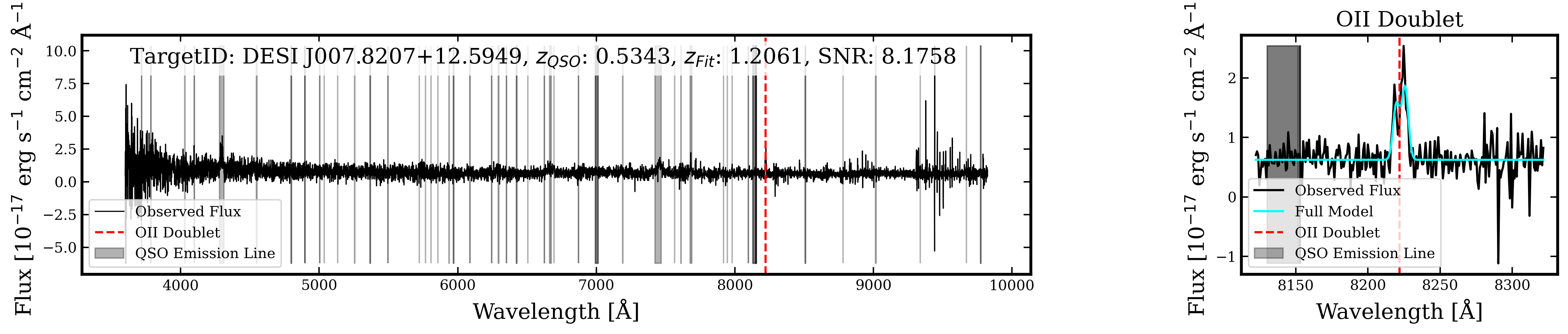}

    \includegraphics[width=\textwidth]{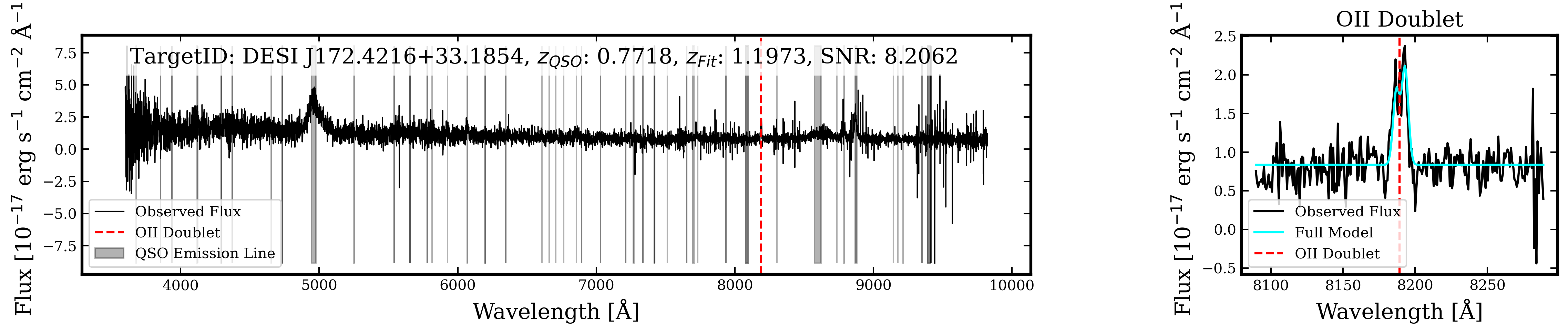}

    
    \includegraphics[width=\textwidth]{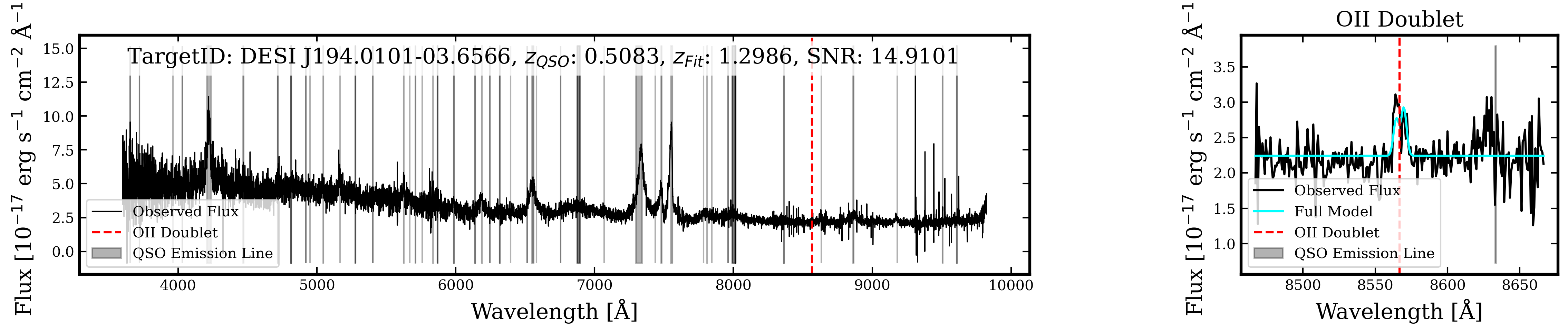}


    \caption{Three A-Grade QSO--ELG lens candidates that only display \Otwo. The left-most plot shows the full spectrum and the rightmost shows the \Otwo doublet. Other strong emission lines such as \Hbeta and \Othreea are not covered by the DESI spectra at these redshifts. The shaded regions are expected QSO emission at that redshift; their widths are determined by the equivalent widths from \citep{SDSS_emission}. The reported SNR correspond to the SNR of the \Otwo emission.}
    \label{fig:SpectraGridO2}
\end{figure*}

\begin{figure*}
    \centering
    \resizebox{\textwidth}{!}{%
        \begin{minipage}{\textwidth}
            \begin{minipage}{0.45\textwidth}
                \centering
                \includegraphics[width=\textwidth]{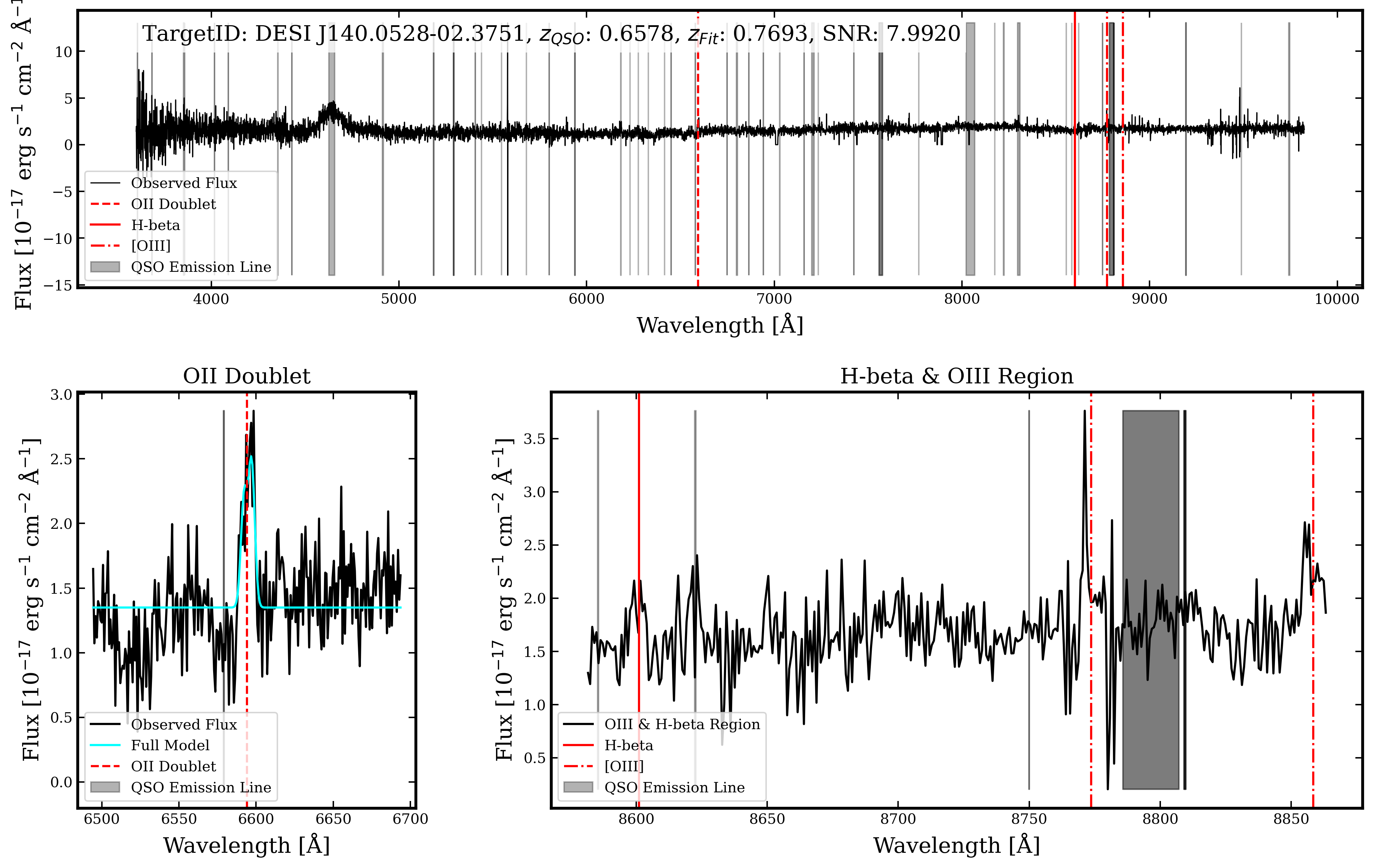}

            \end{minipage}
            \hspace{0.05\textwidth}
            \begin{minipage}{0.45\textwidth}
                \centering
                \includegraphics[width=\textwidth]{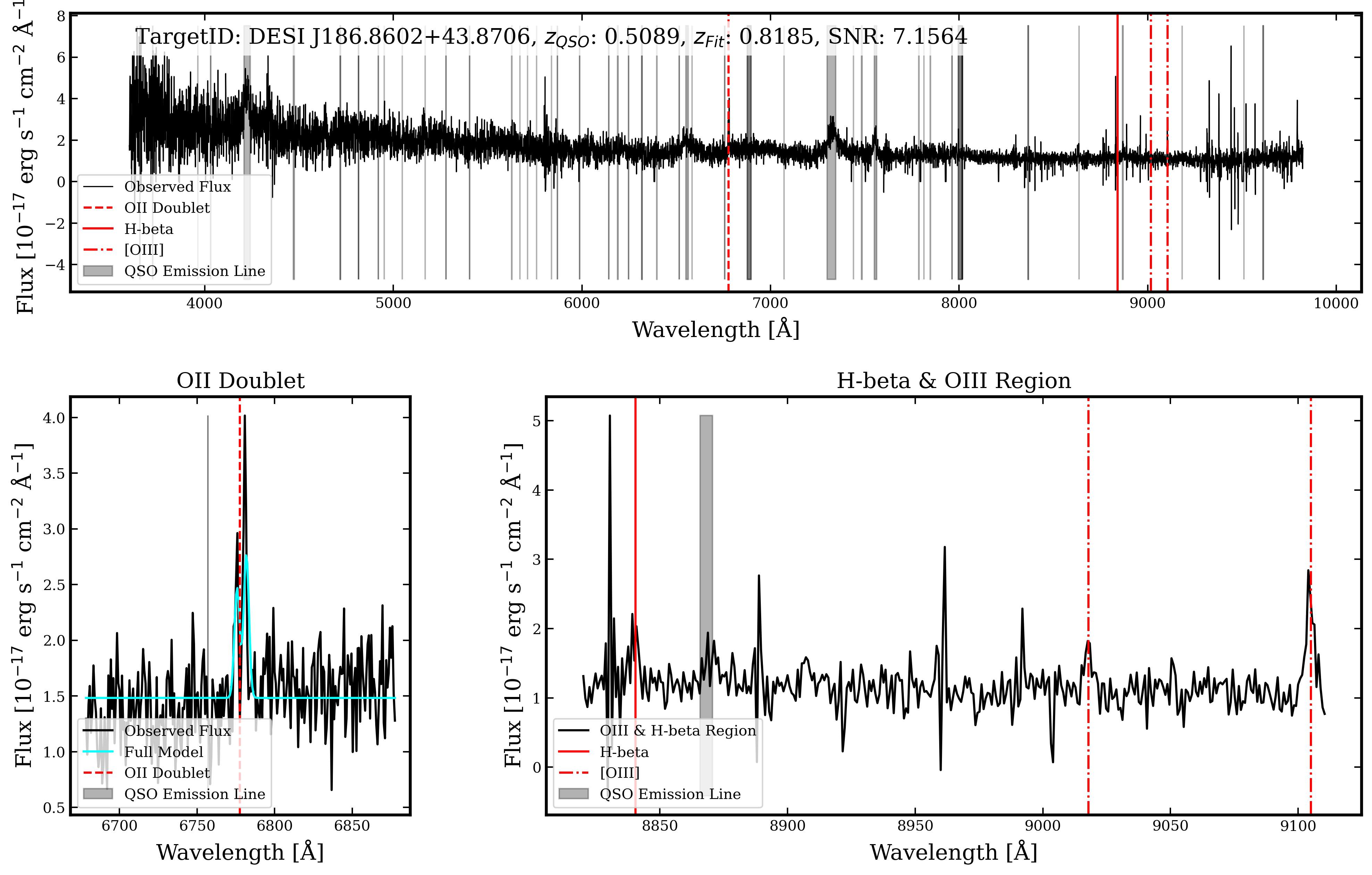}

            \end{minipage}
            
            \vspace{0.2cm} 
            
            \begin{minipage}{0.45\textwidth}
                \centering
                \includegraphics[width=\textwidth]{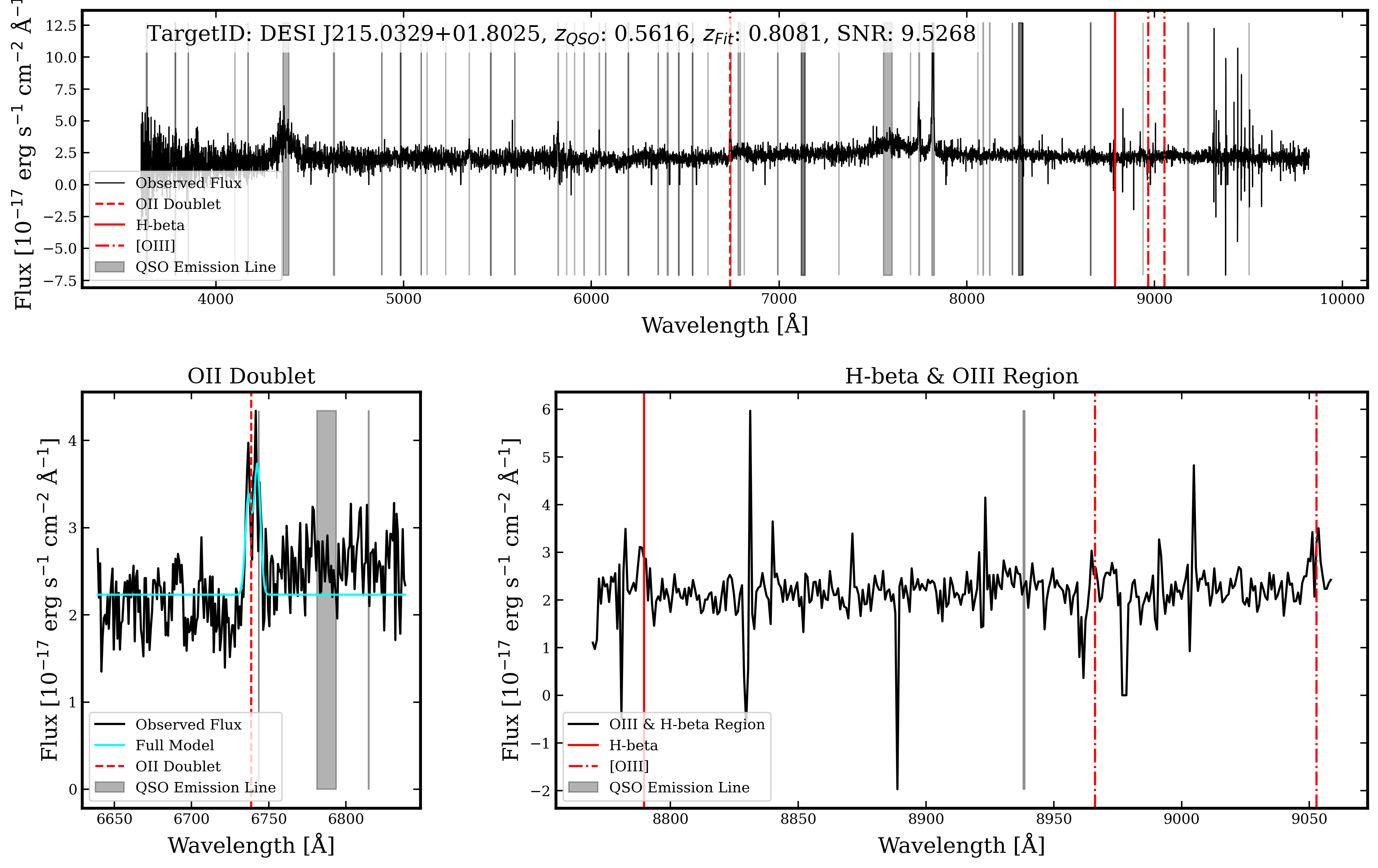}

            \end{minipage}
                \hspace{0.05\textwidth}
            \begin{minipage}{0.45\textwidth}
                \centering
                \includegraphics[width=\textwidth]{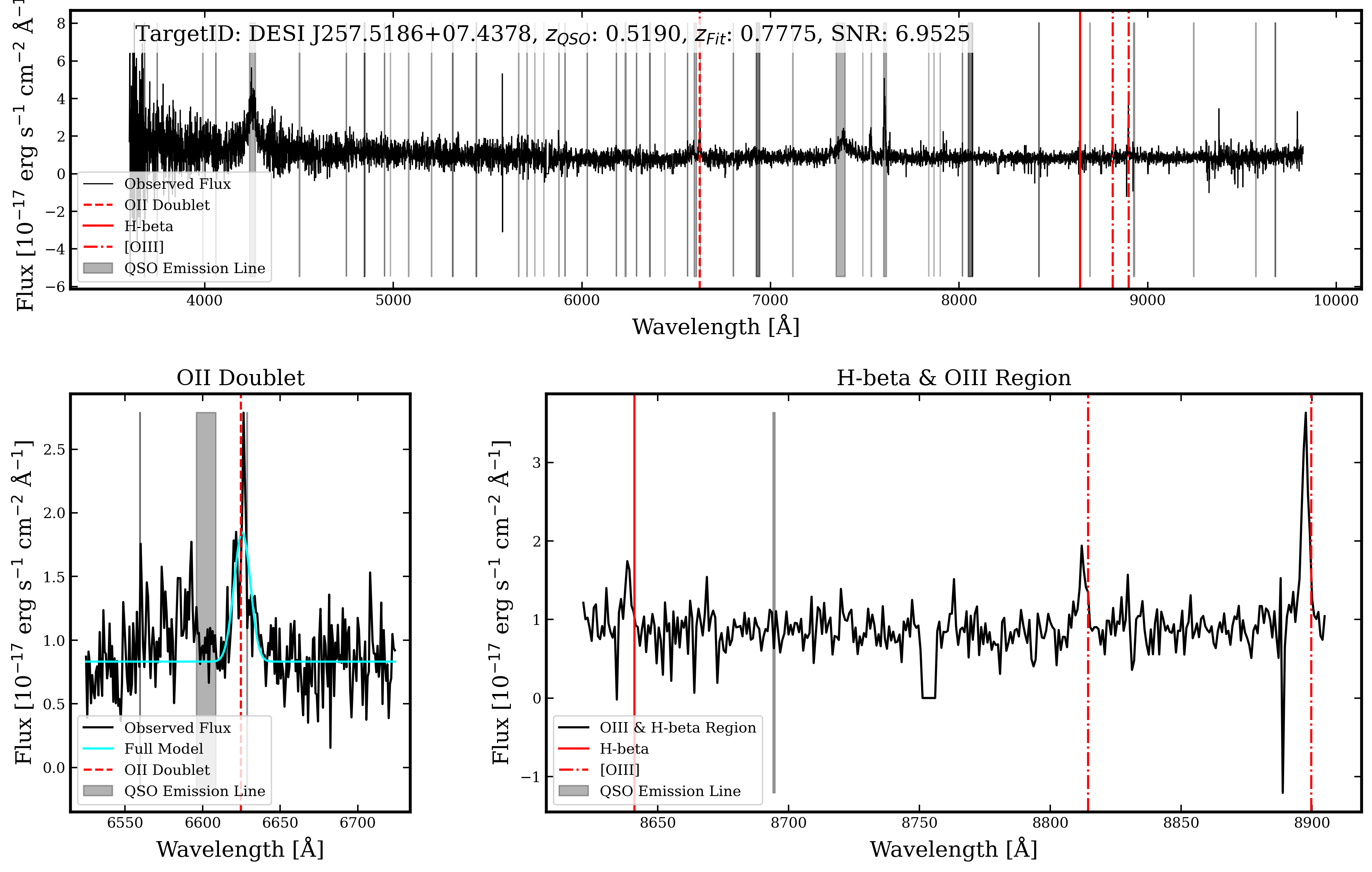}

            \end{minipage}
        \end{minipage}
    }
    \caption{4 QSO-ELG lens candidates that display more than one emission line at a higher redshift than the foreground quasar. The top plot shows the full spectrum, the bottom left shows a zoom-in onto the\Otwo region, and the bottom right displays the \Hbeta and \Othreea (using restframe wavelengths of 4958.911\angstrom and 5006.843 \angstromnospace) emission lines of the higher-redshift source.}
    \label{fig:SpectraGridhbeta}
\end{figure*}

All 7 grade A candidates appear to exhibit strong \Otwo doublet at a higher redshift than the QSO (Figure \ref{fig:SpectraGridO2}). However, four of these candidates also appear to show \Othreea  and H$\beta$ lines at the same redshift. These candidates are DESI J186.8602+43.8706, DESI J140.0528-02.3751, DESI J257.5186+07.4378, and DESI J215.0329+01.8025; spectra for these candidates are shown in Figure \ref{fig:SpectraGridhbeta}.

\begin{figure*}[!h] \centering \includegraphics[width=\textwidth]{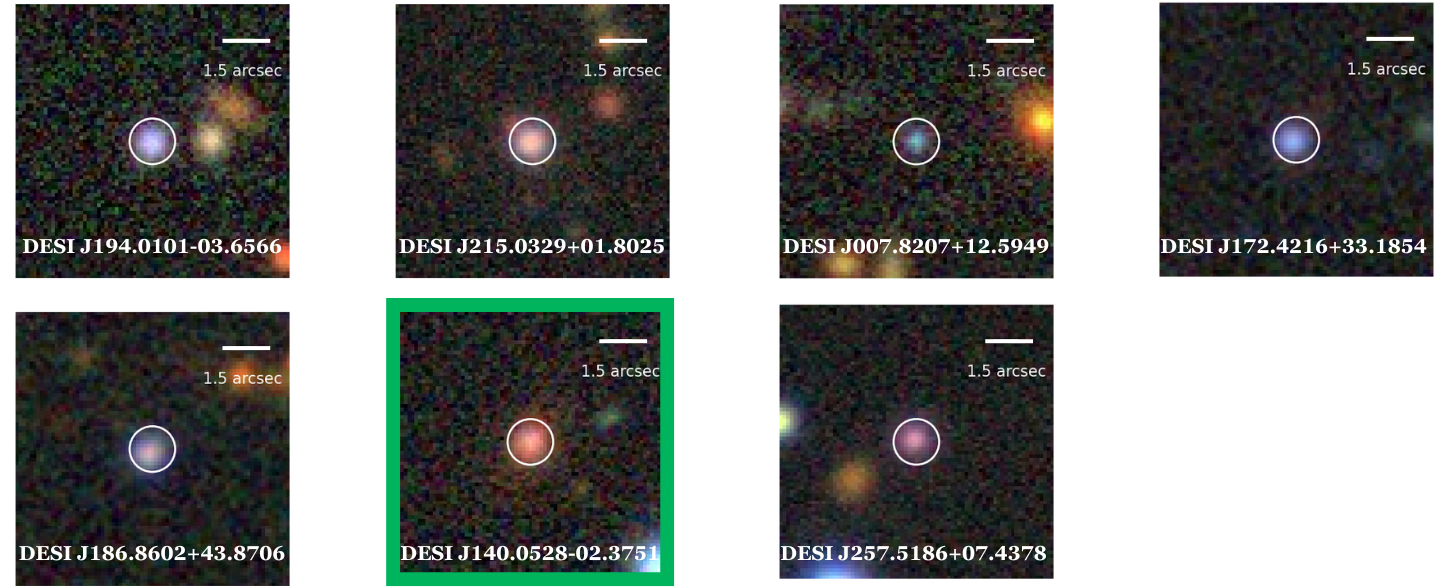} \caption{Legacy Survey images for each of the 7 Candidates. The white circle is $1.5\arcsec$ in diameter, corresponding to the DESI instrument fiber size. The outlined cutout is the recovered candidate from \cite{Meyer_2019}, DESI J140.0528-02.3751.} \label{fig:CUTOUTS} \end{figure*}

In the case of DESI J215.0329+01.8025, while the score of the object is 0.96, we detect a second object within the $1.5\arcsec$ fiber. The DECaLS image of the object in Figure \ref{fig:CUTOUTS} shows the quasar in the center and a red object towards the top left of the target. At the resolution of the DECaLs image it is unclear whether this is a lensed arc or an unlensed nearby galaxy.  If this object is particularly diffuse, its flux may contribute to the additional emission lines observed in the spectra.

DESI J140.0528-02.3751 is the only previously known candidate to be recovered. Not only does its spectrum not seem to be contaminated by nearby galaxies, it was also identified in \cite{Meyer_2019}, using PCA fitting methods. This alignment is encouraging for our CNN classifier approach, as it gave a score of 0.89171, a $z_{\text{fit}}$ of 0.76922, and a $z_\mathrm{ML}$ of 0.85794, as shown in Table \ref{tab:Results} which is consistent with the source redshift estimate of 0.77 from  \citep{Meyer_2019} . This system is the only candidate QSO lens previously found in SDSS that has been observed by DESI so far, and our method was able to recover it. This is also the only one in the DESI DR1 footprint. We note that in Figure \ref{fig:SpectraGridhbeta} the \Hbeta and \Othreea lines appear slightly offset from the apparent emission line location. The observed offset in many of the spectra is due to $z_\mathrm{ML}$ overestimating or underestimating the redshift by $\Delta z$ $\leq 0.001$.

\section{Discussion} 
\label{Discussion}
\subsection{Training and Results} 
\label{Training and Results}
Previous searches for quasars acting as strong gravitational lenses have relied on model-fitting techniques to simultaneously fit the quasar spectrum and the background galaxy spectrum. In contrast, we have opted to use machine learning because of its scalability, particularly when dealing with millions of spectra in larger surveys. To the best of our knowledge, this is the first application of convolutional neural networks to spectroscopic data for gravitational lens searches. In the case of QSO lenses, the diversity of quasar spectra makes template or PCA-based fitting insufficient to fully subtract the quasar flux, often leaving residuals that can be mistaken for background emission lines. A CNN is computationally efficient and can detect anomalies in QSO spectra. We show that our model can generalize the features learned in the training sample to a sample that has never been before. 

Using this methodology we find 7 strong candidates. Our identified candidates predominantly exhibit only the \Otwo emission line, which we attribute to a higher redshift lensed galaxy. This outcome aligns with our expectations, as our training sample was constructed exclusively using ELGs to create the synthetic lensing systems using real data. However, this also introduces several failure modes. First, quasar emission lines at higher redshifts can be misidentified as \Otwo, leading to false positives. Second, prominent skylines, especially at the red end of the spectrum, can be mistakenly classified as \Otwo by the Classification CNN. Given that these generally occur at specific wavelengths, they should be identifiable; however, skylines like the [OH] skyline can be reminiscent of an \Otwo line and, therefore, can pose some challenges for the CNN. Specifically, a number of the contaminants had a redshift estimate of $z=1.49$ due to two especially prominent sky features in the spectra that match the separation of the \Otwo doublet at this redshift. Lastly, artifacts, such as defects in the data, wavelength calibration errors, or mismatch between the calibration of the three arms of the spectrograph, can further confuse the network, either by mimicking real spectral features or distorting the observed wavelengths, leading to misclassification. However, these failure modes are easily identified by visual inspection and larger training sets, and additional quality cuts might also reduce the number of false positives.

\subsection{The Redshift Finder} \label{z finding}
Redshift determination is challenging, and careful consideration is required when defining a redshift-finding procedure. Typically, using a CNN to determine the redshift is not ideal due to the spread around the true value, which often results in an imprecise estimate. However, in our case, we only require an estimate of the redshift. We perform a local fit of a double Gaussian to model the doublet in a $\Delta z=$0.1 redshift range around the redshift found from the regression network. At high and medium SNR, this Gaussian fit led to more precise estimation of the redshift. However, we notice that the Gaussian fits degrade the performance in low SNR regimes, especially when the \Otwo line is observed in the near infrared. This is due to the proximity of sky lines in this part of the spectrum, which are sometimes mistaken for the \Otwo lines during the Gaussian fit, leading to inaccurate measurement. A proper masking of these sky lines would improve our performances in the low SNR regime. 


Our redshift finder plays a critical role in calculating the SNR. The calculated SNR is dependent on the redshift of the object, as found by both the Gaussian model and the redshift finder. If an object has an SNR below 3 as determined by the redshift finding procedure, we exclude it from our primary results. In doing so, we may be at risk of removing genuine candidates due to an incorrect redshift prediction. Nonetheless, the results found from both the classifier and the regression analysis demonstrate that machine learning can learn the complexities of quasar emission and only isolate the feature of interest. 



\section{Summary} 
\label{Summary}
We have developed a data-driven pipeline to identify quasars acting as strong gravitational lenses in spectroscopic data from DESI DR1, which span the redshift range $0.03 \leq z \leq 1.8$. Our method leverages convolutional neural networks trained on realistic mock lenses constructed from real DESI QSO and ELG spectra. Applying this pipeline to 812,118 quasars, we identify seven high-quality lens candidates, including one previously reported system \citep[DESI J140.0528-02.3751;][]{Meyer_2019}.

Our search strategy includes generating mock blended spectra, training a CNN classifier in two phases to distinguish lensing features, and applying a CNN-based redshift estimator to recover the background source redshift ($z_{\rm ELG}$). The final candidates are selected after visual inspection and identification of the background emission lines. Given the small expected Einstein radii and the brightness of the foreground quasars, such features are unlikely to be visible from the ground. To confirm the lensing nature of these systems, high-resolution imaging with HST or Euclid will be required.

Our discovery of new quasar lens candidates opens the door to assembling the first statistical sample of QSOs acting as strong lenses and demonstrates that data-driven methods can uncover these rare systems at scale. The resulting sample will provide a powerful new avenue to study the coevolution of black holes and galaxies, anchoring scaling relations with direct mass measurements across cosmic time.

\section*{Code Availability}
The Pipeline and associated code to recreate plots are freely available \href{https://doi.org/10.5281/zenodo.17457851}{here}.

\section*{Acknowledgments}
This work was initiated as a part of the postbaccalaureate program at the Kavli Institute for Particle Astrophysics and Cosmology, and also received support from the U.S. Department of Energy under contract number DE-AC02-76SF00515 to SLAC National Accelerator Laboratory. The authors thank Jelle Aalbers, Phil Marshall, Padma Venkatraman, and Sydney Erickson and the GFC group at Stanford for helpful discussions at the early stages of this project. MM acknowledges support by the SNSF (Swiss National Science Foundation) through mobility grant P500PT\_203114 and return CH grant P5R5PT\_225598.

M.S. acknowledges support by the State Research Agency of the Spanish
Ministry of Science and Innovation under the grants 'Galaxy Evolution
with Artificial Intelligence' (PGC2018-100852-A-I00) and 'BASALT'
(PID2021-126838NB-I00) and the Polish National Agency for Academic
Exchange (Bekker grant BPN/BEK/2021/1/00298/DEC/1). This work was
partially supported by the European Union's Horizon 2020 Research and
Innovation program under the Maria Sklodowska-Curie grant agreement (No.
754510).

This material is based upon work supported by the U.S. Department of Energy (DOE), Office of Science, Office of High-Energy Physics, under Contract No. DE–AC02–05CH11231, and by the National Energy Research Scientific Computing Center, a DOE Office of Science User Facility under the same contract. Additional support for DESI was provided by the U.S. National Science Foundation (NSF), Division of Astronomical Sciences under Contract No. AST-0950945 to the NSF’s National Optical-Infrared Astronomy Research Laboratory; the Science and Technology Facilities Council of the United Kingdom; the Gordon and Betty Moore Foundation; the Heising-Simons Foundation; the French Alternative Energies and Atomic Energy Commission (CEA); the National Council of Humanities, Science and Technology of Mexico (CONAHCYT); the Ministry of Science, Innovation and Universities of Spain (MICIU/AEI/10.13039/501100011033), and by the DESI Member Institutions: \url{https://www.desi.lbl.gov/collaborating-institutions}.

The DESI Legacy Imaging Surveys consist of three individual and complementary projects: the Dark Energy Camera Legacy Survey (DECaLS), the Beijing-Arizona Sky Survey (BASS), and the Mayall z-band Legacy Survey (MzLS). DECaLS, BASS and MzLS together include data obtained, respectively, at the Blanco telescope, Cerro Tololo Inter-American Observatory, NSF’s NOIRLab; the Bok telescope, Steward Observatory, University of Arizona; and the Mayall telescope, Kitt Peak National Observatory, NOIRLab. NOIRLab is operated by the Association of Universities for Research in Astronomy (AURA) under a cooperative agreement with the National Science Foundation. Pipeline processing and analyses of the data were supported by NOIRLab and the Lawrence Berkeley National Laboratory. Legacy Surveys also uses data products from the Near-Earth Object Wide-field Infrared Survey Explorer (NEOWISE), a project of the Jet Propulsion Laboratory/California Institute of Technology, funded by the National Aeronautics and Space Administration. Legacy Surveys was supported by: the Director, Office of Science, Office of High Energy Physics of the U.S. Department of Energy; the National Energy Research Scientific Computing Center, a DOE Office of Science User Facility; the U.S. National Science Foundation, Division of Astronomical Sciences; the National Astronomical Observatories of China, the Chinese Academy of Sciences and the Chinese National Natural Science Foundation. LBNL is managed by the Regents of the University of California under contract to the U.S. Department of Energy. The complete acknowledgments can be found at \url{https://www.legacysurvey.org/}.

Any opinions, findings, and conclusions or recommendations expressed in this material are those of the author(s) and do not necessarily reflect the views of the U. S. National Science Foundation, the U. S. Department of Energy, or any of the listed funding agencies.

The authors are honored to be permitted to conduct scientific research on I'oligam Du'ag (Kitt Peak), a mountain with particular significance to the Tohono O’odham Nation. 

\vspace{5mm}
\software{astropy \citep{2013A&A...558A..33A,2018AJ....156..123A,2022ApJ...935..167A}, Numpy \citep{2020Natur.585..357H}, Matplotlib \citep{2007CSE.....9...90H}, SciPy \citep{2020NatMe..17..261V}, TensorFlow \cite{tensorflow2015-whitepaper}, SciKit \cite{scikit-learn}, pandas \citep{reback2020pandas} \citep{mckinney-proc-scipy-2010}}

\clearpage
\appendix

\section{Grade B Candidates}
\label{Sec: GradeB}
Below we present the candidates assigned a grade B, as it was unclear whether the detected feature correspond to the \Otwo\ emission lines. We present the spectra below the resulting table in the same fashion as Figure \ref{fig:SpectraGridO2}.

\begin{figure}[hbt]
    \centering
    \resizebox{\textwidth}{!}{%
        \begin{minipage}{\textwidth}
            \begin{minipage}{\textwidth}
                \centering
                \includegraphics[width=\textwidth]{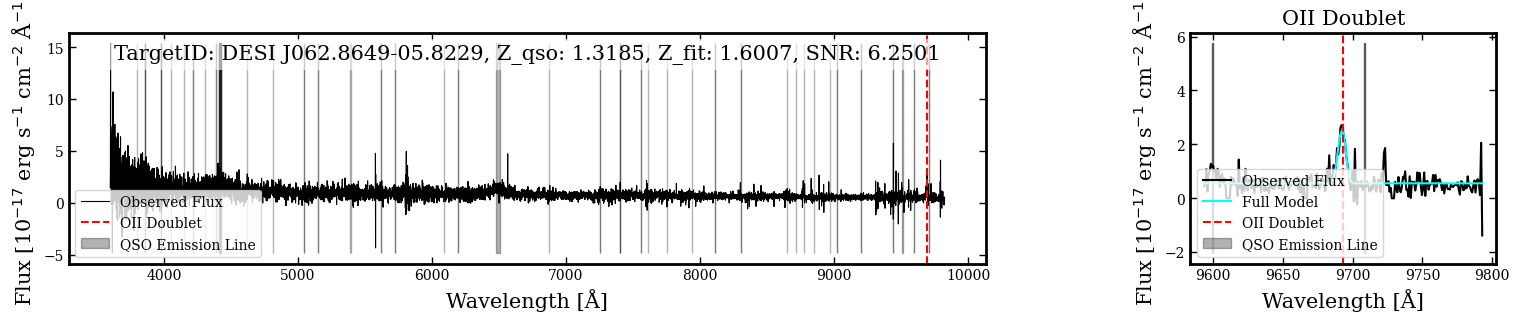}
            \end{minipage}
            
            \vspace{0.2cm} 

            \begin{minipage}{\textwidth}
                \centering
                \includegraphics[width=\textwidth]{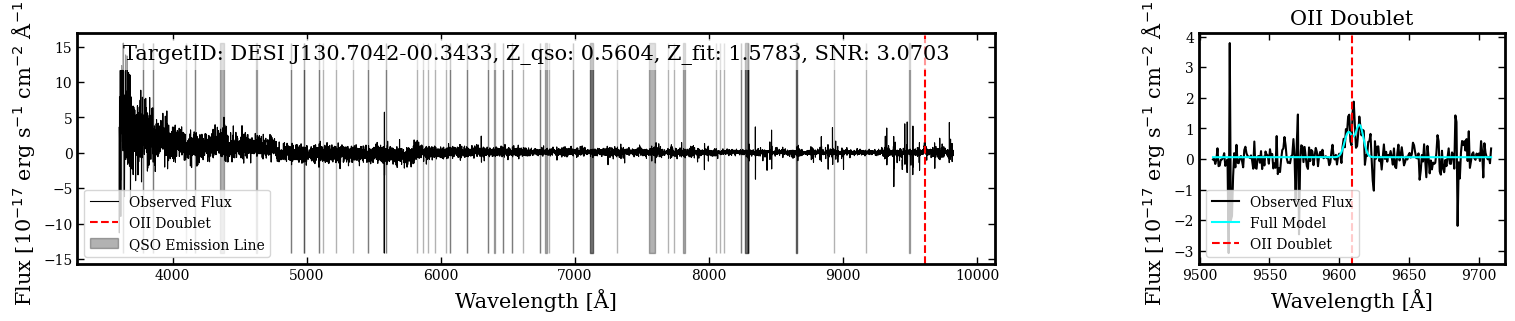}
            \end{minipage}

        \end{minipage}
    }
    \caption{2 with the the lowest SNR B-Grade QSO-ELG candidates with a possible \Otwo emission present. The top plots show the full QSO spectrum, where the gray regions are the equivalent width of the QSO emission with respect to redshift, and the red lines are the background galaxy emissions.}
    \label{fig:GradeBOII}
\end{figure}

\begin{figure}[hbt]
    \centering
    \resizebox{\textwidth}{!}{%
        \begin{minipage}{\textwidth}

            \begin{minipage}{\textwidth}
                \centering
                \includegraphics[width=\textwidth]{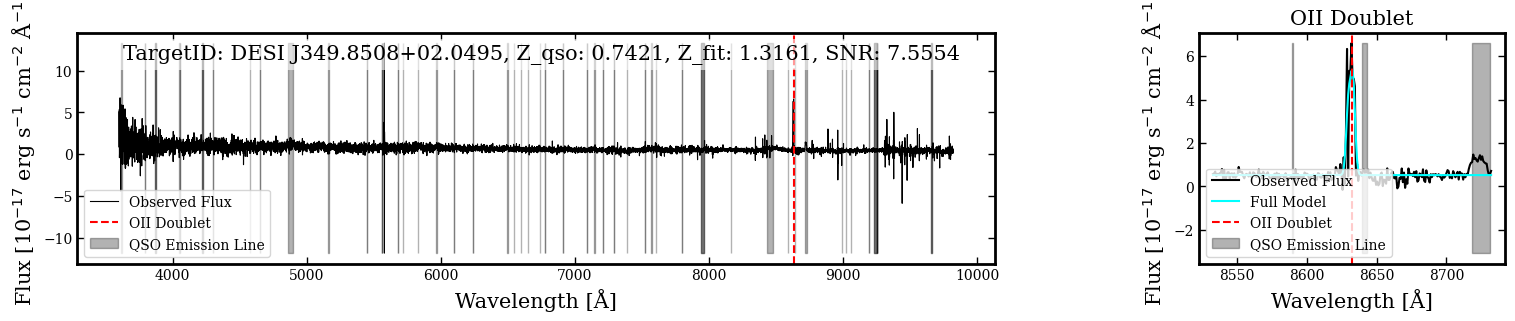}
            \end{minipage}

            \begin{minipage}{\textwidth}
                \centering
                \includegraphics[width=\textwidth]{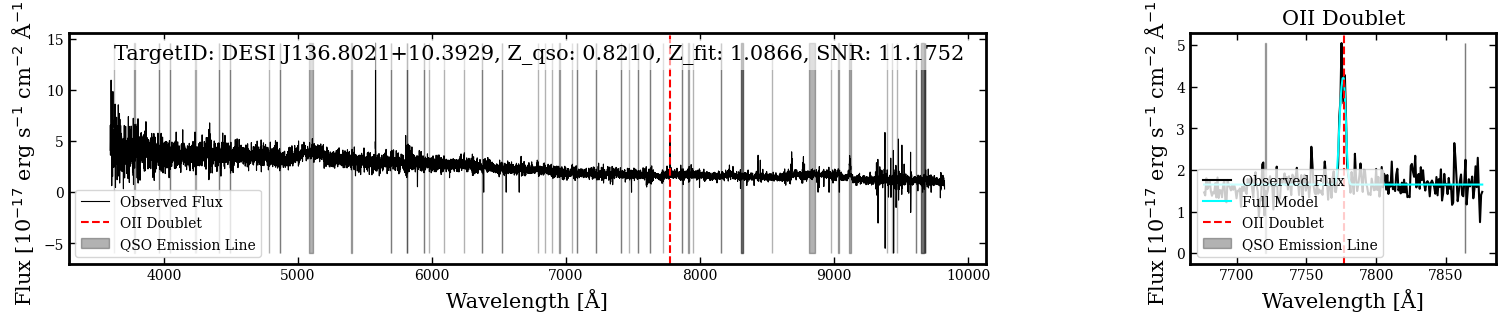}
            \end{minipage}
        \end{minipage}
    }
    \caption{2 highest SNR B-Grade QSO-ELG candidates with a possible \Otwo emission present.}
    \label{fig:GradeB_part2}
\end{figure}

\begin{table*}[!htp]\centering
\scriptsize
\begin{tabular*}{\textwidth}{@{\extracolsep{\fill}} lrrrrrrr @{}}
\toprule
Name & TargetID & Z$_{QSO}$ & Score & Z$_{ML}$ & Z$_{Fit}$ & $\chi^2$ & SNR [OII] \\
\midrule
DESI J062.8649-05.8229 & 39627647021484584 & 1.3185 & 0.9941 & 1.5477 & 1.6008 $\pm$ 0.0013 & 1.0004 & 6.2501 \\
DESI J349.8508+02.0495 & 39627838915086574 & 0.7421 & 0.7803 & 1.3002 & 1.3161 $\pm$ 0.0001 & 2.0294 & 7.5554 \\
DESI J130.7042-00.3433 & 39627780878506428 & 0.5604 & 0.9723 & 1.5675 & 1.5783 $\pm$ 0.0008 & 1.1548 & 3.0703 \\
DESI J136.8021+10.3929 & 39628039482511257 & 0.8210 & 0.7598 & 1.0630 & 1.0866 $\pm$ 0.0002 & 1.0062 & 11.1752 \\
\bottomrule
\end{tabular*}
\caption{Grade B QSO–ELG candidates found in DESI DR1 from the full pipeline. Column names are the same as those found in Table \ref{tab:Results}.}
\label{Tab:GradeB}
\end{table*}

\clearpage
\bibliography{QSOPaperRef}
\bibliographystyle{aasjournal}

\end{document}